\documentclass[a4paper,aps,pra,notitlepage,superscriptaddress,nofootinbib,reprint]{revtex4-1}

\usepackage{graphicx}
\usepackage{graphics}
\usepackage{color}
\usepackage{amsmath,amssymb,amsthm}
\usepackage[usenames,dvipsnames]{xcolor}
\usepackage[right=1.5cm, left=2cm, top=1.5cm, bottom=1.5cm]{geometry}
\newcommand{\be}{\begin{equation}}
\newcommand{\ee}[1]{\label{#1} \end{equation}}

\def\f12{\frac{1}{2}}

\usepackage{comment} 
\usepackage[normalem]{ulem}
\usepackage{cancel} 

\begin{document}

\title{\fontsize{0.5cm}{0.5em}\selectfont Comment on ``Quantum mechanics for non-inertial observers'' }

\author{Igor Pikovski}
\affiliation{ITAMP, Harvard-Smithsonian Center for Astrophysics, Cambridge, MA 02138, USA}
\affiliation{Department of Physics, Harvard University, Cambridge, MA 02138, USA}
\author{Magdalena Zych}
\author{Fabio Costa}
\affiliation{Centre for Engineered Quantum Systems, School of Mathematics and Physics, The University of Queensland, St Lucia, QLD 4072, Australia }
\author{\v{C}aslav Brukner}
\affiliation{Vienna Center for Quantum Science and Technology (VCQ), University of Vienna, Faculty of Physics, Boltzmanngasse 5, A-1090 Vienna, Austria}
\affiliation{Institute for Quantum Optics and Quantum Information (IQOQI), Austrian Academy of Sciences, Boltzmanngasse 3, A-1090 Vienna, Austria}

\date{\today}
\begin{abstract}
In a recent paper (\href{http://arxiv.org/abs/arXiv:1701.04298}{{\ttfamily
  arXiv:1701.04298 [quant-ph]}}) Toro{\v{s}}, Gro{\ss}ardt and Bassi claim that the potential necessary to support a composite particle in a gravitational field must necessarily cancel the relativistic coupling between internal and external degrees of freedom. As such a coupling is responsible for the gravitational redshift measured in numerous experiments, the above statement is clearly incorrect. We identify the simple mistake {in the paper} responsible for the incorrect claim.
\end{abstract}

\maketitle

According to General Relativity, the internal energies of composite systems appear shifted depending on their velocity and position in a gravitational field. This effect can be described in dynamical terms: Let $H_0$ be the internal Hamiltonian of a composite system, as described in its rest frame, and let $x$ and $p$ be its position above Earth and momentum, respectively. Then, in a low-energy approximation and neglecting the $O(c^{-2})$ corrections to position and momentum that are irrelevant to the argument, the dynamics is described by the Hamiltonian:
\begin{equation}
H = \frac{p^2}{2m}+ mgx  +  \left(1-\frac{p^2}{2m^2c^2} + \frac{gx}{c^2}\right) H_0 +U_{\textrm{ext}}(x).
\label{ham}
\end{equation}
Here, $U_{\textrm{ext}}(x)$ is an external potential, which is necessary in general to constrain the particle along a non-inertial world-line. In a quantum regime, the same Hamiltonian produces entanglement between the external and internal degrees of freedom of the particle~\cite{zych_quantum_2011, pikovskiuniversal2015, Zych2016, bushev2016single, Krause2016, Orlando2016, Pikovski2017}.

A simple example is a particle kept at a fixed height $x_0$: the above expression implies that the internal dynamics is driven by the shifted Hamiltonian $(1 + \frac{gx_0}{c^2})H_0$, with respect to a particle at height $x=0$. Known as gravitational time dilation, this effect can be measured by comparing the ticking rates  of two clocks at different heights. Recent experiments have measured the effect down to a height difference of less than $1$ meter \cite{chou2010optical}.

In their recent paper \cite{torovs2017quantum}, Toro{\v{s}}, Gro{\ss}ardt, and Bassi re-derive the above Hamiltonian, in agreement with Refs.~\cite{zych_quantum_2011, pikovskiuniversal2015, Zych2016, bushev2016single, Krause2016, Orlando2016, Pikovski2017}. However, the authors come to the surprising conclusion that, in order to support the clocks, the external potential must necessarily cancel all position-dependent couplings in the Hamiltonian \eqref{ham}. The claim is that the only potential capable of keeping a particle at a fixed height is $U_{\textrm{ext}}(x)= - (m+  H_0/c^2)g x$. If this were the case, the clocks used in experiments---which are indeed kept at fixed heights---would be observed to tick at equal rates, regardless of their position, because the term $g x H_0/c^2$ would effectively vanish from the Hamiltonian. Such a term is however observed in experiments.

The issue can be identified in section IV of Ref.~\cite{torovs2017quantum}, where it is imposed that, for a particle at rest, momentum must be constant. Then, Hamilton equations of motion imply $\dot{p}=-\partial H/\partial x =0$ and it is concluded that $H$ cannot depend on $x$. This reasoning is the mistake: To trap a particle, the above quantities only need to vanish for a \emph{specific solution}, not identically in the equations of motion. For example, for a potential with a minimum at $x_0$,  the constant solution $x(t)=x_0$ has constant momentum, i.e. $\left.\dot{p}\right|_{x=x_0}=\left.-\partial H/\partial x\right|_{x=x_0}=0$. For this to hold, $U_{\textrm{ext}}$ does not need to depend on $H_0$, and in fact never does in actual experiments probing time dilation \cite{Pound1959,Hafele1972,chou2010optical}. A similar issue remains for the quantum version of the argument: the authors of Ref.~\cite{torovs2017quantum}  require that acceleration vanishes for the expectation value of position and conclude that $U_{\textrm{ext}}(x)$ must exactly cancel the gravitational terms.
To the contrary, typical trapping potentials would have bound eigenstates as solutions (e.g.~directly measured with ``bouncing neutrons'' \cite{nesvizhevsky2002quantum}), for which the average position \emph{is} constant and without cancelling the gravitational time dilation terms.

The total Hamiltonian found by Toro{\v{s}} and colleagues is in fact that of a free particle in Minkowski coordinates. This is not surprising: they consider a particle in Rindler coordinates (homogeneous gravity) and find that a potential that identically cancels all effects of uniform acceleration must couple to the total mass-energy of the system.
The only interaction in nature that can universally do that (regardless of the nature of the internal structure) is gravity itself---their $U_{\textrm{ext}}$   indeed exactly mimics a homogeneous gravitational potential. The equivalence principle states that, locally, gravity can be canceled by moving to an accelerated frame. The situation in Ref.~\cite{torovs2017quantum} can thus be seen as just the converse: the effects observed in an accelerated frame are canceled identically by an appropriate gravitational potential.

As a last remark, once a world line of a system is specified, its proper time does not depend on the external potentials -- a core aspect of general relativity. Indeed, such potentials are typically not necessary to predict the effects of time dilation in experiments~\cite{hafele1972predicted}.

In conclusion, contrary to the claim of Toro{\v{s}}, Gro{\ss}ardt, and Bassi, a potential that keeps a particle from falling does not in general couple to the internal energy and thus does not cancel the relativistic coupling between internal and external degrees of freedom -- as confirmed by experiments. Since it is the same coupling that produces the effects in Refs.~\cite{zych_quantum_2011, pikovskiuniversal2015, Krause2016, Orlando2016,  Zych2016, bushev2016single, Pikovski2017}, such effects, and in particular time dilation induced decoherence~\cite{pikovskiuniversal2015}, would not be cancelled by external potentials either.

\bibliographystyle{linksen}
\bibliography{bibliocomment}
\end{document}